# Comparison of measured and simulated spin-wave mode spectra of magnetic nanostructures

H. T. Nembach[1,2], R.D. McMichael[3], M.L. Schneider[2], J.M. Shaw[2], T.J. Silva[2]

[1]JILA, University of Colorado, Boulder, Colorado 80309, USA

[2]Quantum Electromagnetics Division, National Institute of Standards and Technology, Boulder, Colorado 80305, USA

[3] Nanoscale Device Characterization Division, National Institute of Standards and Technology, Gaithersburg, Maryland 20899, USA

Motivated by the importance of magnetization dynamics in nanomagnets for the development and optimization of magnetic devices and sensors, we measured and modeled spin wave spectra in patterned elliptical nanomagnets. Ferromagnetic resonance spectra for multiple nanomagnets of $Ni_{80}Fe_{20}$, fabricated by electron-beam lithography to have nominal short-axes of 200 nm or 100 nm, were measured by use of heterodyne magneto-optical microwave microscopy. Scanning electron microscope images taken of the same nanomagnets were used to define element shapes for micromagnetic simulations. Measured spectra show significant differences between nominally identical nanomagnets, which could be only partially attributed to uncontrolled shape variations in the patterning process, as evidence by the limited agreement between the measured and simulated spectra. Agreement between measurements and simulations was improved by including a zone of reduced magnetization and exchange at the edges of the nanomagnets in the simulations. Our results show that the reduction of shape variations between individual magnetic random-access memory elements can potentially improve their performance. However, unambiguous determination of materials parameters in nanomagnets based on analysis and modeling of spin wave spectra remains problematic.

Nanomagnets are the building blocks of hard disk drive read heads [1], magnetic random access memory (MRAM) [2–4], and have promise for emerging applications in probabilistic [5–7] and neuromorphic computing [8–10]. These applications all require the understanding and exploitation of high-speed dynamics within the nanomagnets. MRAM has been used in specialized hardware for several years but is poised to enter a much broader range of applications as embedded MRAM. Some of the critical parameters for MRAM devices include write error rate and thermal stability. The write error rate is the rate of failed switching events for a given number of attempts. High write error rates require more computation overhead with respect to error correction, which will tend to drive up the overall cost. Thermal stability determines the data retention time and is proportional to the energy barrier, which prevents undesired thermal fluctuations between the two magnetic states. A high energy barrier provides high thermal stability though typically at the expense of larger currents required for write operations[4]. A detailed understanding of the switching process is critical to guide the design of MRAM.[11] The switching process depends on the physical switching mechanism in conjunction with both the material parameters and the shape of the MRAM cell.

We investigate the role of shape imperfections on the mode dynamics in single nanomagnets. As is known, shape imperfections give rise to spatially varying demagnetizing fields. Such local variation in the demagnetization fields can lead to nucleation sites for the switching[12]. It has been demonstrated that higher write error rates are correlated with the existence of more resonances in spin-torque ferromagnetic resonance (ST-FMR) measurements of devices[13]. These additional resonances can be the result of deviations from the ideal shape of the device.

In addition to shape imperfections, non-uniformity in materials parameters across a device can also be introduced during the patterning process. For example, transmission electron microscopy images have found a ring of about 12 nm to 20 nm of altered contrast at the edges of $(Co/Pd)_n$ multilayer nanomagnets[14]. It has also been shown that the stiffness fields, and consequently the resonance fields, for localized spinwave modes in $Ni_{80}Fe_{20}$ stripes depend on the sidewall angle[15]. This suggests that the details of spin-wave modes near patterned edges are highly susceptible to both edge damage and unintended shape variations. Nembach *et al*. demonstrated that the number of localized modes could be increased by intentionally distorting the elliptical shape when measuring large ensembles of nanomagnets by use of low-resolution Brillouin light scattering spectroscopy [16]. However, ensemble measurements are prone to interpretation difficulties when nanomagnet variations cause significant line broadening[17]. Measurements of individual nanomagnets are required to fully understand how much variation in the localized spin-wave spectra for nominally identical structures actually exists.

The heterodyne magneto-optical Kerr effect microscope (H-MOMM) is an optical tool for the study the spin-wave spectra of well-spaced individual nanomagnets[18]. H-MOMM was previously used to investigate damping enhancement in patterned nanomagnets as small as 100 nm, where the spatial curvature of a given spin-mode had a direct impact on the frequency dependence of the measured linewidth[19]. In the present study, we apply H-MOMM to survey the spin-wave spectra of nominally identically nanomagnets in sparse patterned arrays. The spectra are surprisingly variable, given the tolerances of the lithographic processing employed. It appears as if each nanomagnet has its own spectroscopic fingerprint that makes it uniquely identifiable in spite of the process control. To understand how the spin-wave spectra can vary so much, we used scanning electron microscopy (SEM) as a dimensioning tool to image the same nanomagnets that were used for the H-MOMM measurements, and the dimensions obtained were then input to micromagnetic simulations. We obtained only qualitative agreement between the H-MOMM measurements and micromagnetics simulations, which implies that more subtle details associated with edge structure are strongly affecting the boundary conditions for the localized spin-wave modes. As an example of such an edge detail that would not be apparent in SEM micrographs, we investigate how a simple gradation of magnetization near the edges of the nanomagnets can significantly alter the simulated spectra. Our results confirm qualitatively that such gradients, in conjunction with shape variations, are sufficient to cause the variations in the spectra. This highlights the need for additional metrological tools to address how lithographic patterning affects the internal energetic landscape of patterned nanomagnets.

In this work, we prepared two sets of $Ni_{80}Fe_{20}$ elliptical nanomagnets with nominal long axis lengths of 240 nm and 120 nm and short axis of 200 nm and 100 nm: Thin-film layers of 3 nm Ta/10 nm $Ni_{80}Fe_{20}$/5 nm $Si_3N_4$ were dc-magnetron sputtered onto a sapphire substrate before a 15 nm diamond-like carbon (DLC) layer was deposited via ion-beam deposition in a separate vacuum chamber. Electron beam lithography was then used to expose a 100 nm thick polymethyl methacrylate layer which was developed in an methyl isobutyl ketone : isopropanol solution. A 5 nm Cr layer was then deposited and lifted off via ion beam deposition. The pattern was transferred to the DLC layer via an $O_2$ plasma etch before the Cr was removed with a wet etch. The final pattern transfer to the $Ni_{80}Fe_{20}$ layer was accomplished by a 300 eV Ar ion mill. Finally, the remaining DLC was removed by a second $O_2$ plasma etch. The ellipses were imaged with an SEM as shown in the insets in the left columns of Fig. 1 and Fig. 2. Images of additional nanomagnets are shown in the Supplemental Information.

The spin wave mode spectra of the magnetization dynamics were measured with a heterodyne magneto-optical microwave microscope (H-MOMM), see Fig. 3 for a simplified sketch of the setup[18]. The H-MOMM employs two frequency tunable single frequency lasers operating at 532 nm. Part of both laser beams is split-off and focused onto a high-speed photodiode, where they generate microwaves at the difference frequency of both laser beams. The microwaves are amplified and fed into a coplanar waveguide, with the sample located on top. One of the laser beams is focused with an objective lens with a NA = 0.9 onto the sample. When the external magnetic field meets the resonance condition, the precessing magnetization modulates the polarization of the back-reflected light due to the magneto-optical Kerr effect. The back-reflected light is then mixed with the other laser beam at a beam splitter cube. The two out-going beams are finally focused onto a differential detector. As a result of the demodulation of the AC signal achieved by remixing the two laser beams, the measured DC-signal on the differential detector is proportional to the amplitude of the Kerr rotation generated by the precessing magnetization of the back-reflected laser beam. The H-MOMM allows to measure nanomagnets fabricated from individual magnetic layers, which are the building blocks of magnetic logic and memory devise. This in contrast to electrical measurements, which can be for example based on ST-FMR[20,21]. These measurements require at least two magnetic layers, the reference and the free layer. In general, ST-FMR measurements probe the performance of the complete device and as such provide complementary information to H-MOMM measurements on the building blocks. The impact of process induced edge damage on spin-wave modes has been successfully measured with ST-FMR[13,22].

Representative spectra for the 200 nm and 100 nm nanomagnets are shown in Fig. 1 and Fig. 2 respectively. The magnetic field was applied along the long axis of the nanomagnets and the excitation frequency was ~8.2 GHz. Most of the spectra for the 200 nm nanomagnets include one intense peak and two weaker peaks. Two peaks are visible in most of the 100 nm nanomagnets' spectra. The nanomagnets within each of these sets all have nominally the same dimensions. Spectra for other, nominally identical nanomagnets are provided in the Supplemental Information.

We carried out micromagnetic simulations using the Object Oriented MicroMagnetic Framework (OommF)[23]. To determine the shape for modeled nanomagnets, greyscale SEM images of the nanomagnets were converted into binary images using a thresholding algorithm. The original SEM images, see insets of Figs.1 (a,c) and 2 (a,c), were given a Gaussian blur over 1.4 nm (3 pixels), rescaled by 25 % and given a secondary blur over 3.8 nm. A threshold value was determined using Otsu's method[24]. The resulting sample boundaries for the respective nanomagnets are shown in the insets of Figs. 1 (b,d) and 2 (b,d).

The simulated spectra were extracted from impulse response calculations made at an array of applied field values in the experimental range. The modeling also provides the spatial profile of the spin wave modes, see the insets of Fig. 1 (b) and (d) and Fig 2 (b) and (d). The spatial profiles show that the strongest peak for the larger nanomagnet is mainly localized in the center of the nanomagnet (center-mode), whereas the two weaker modes are localized at the ends of the nanomagnet (end-modes). For the smaller nanomagnets, only end-modes are visible in the experimental range, but the model also produces a center-mode at higher frequencies.

In the larger magnets, one of the most striking differences between measured and modeled spectra is that the resonance field difference between the center- and the end-modes is much larger in the simulations. We ran additional simulations to check the possibility of edge damage, which can include angled sidewalls and reduced saturation magnetization may be responsible for these differences[15]. In order to model damage as a zone of reduced magnetization at the edges, we started out with ideal ellipses with dimensions 262 nm x 190 nm and 126 nm x 90 nm, which were subdivided into 2 x 2 x 10 $nm^3$ cells. We then reduced the magnetization at the edge by averaging the magnetization for each cell over a disk with a set radius $r$, where $r$ ranged from 2 nm to 8 nm. We held the exchange length constant for the area of reduced magnetization, which is equivalent to assuming that the exchange stiffness $A$ scales as $M_s^2$.

The results of these modified-edge simulations are shown in Fig. 4. Panels (a) and (b) show spectra calculated for larger and smaller ellipse sizes respectively, with increasing edge damage from top to bottom as indicated in the center images. In Fig. 4 (a), the effect of the reduced magnetization and exchange at the edges can be seen in the simulated spectra as a slight increase of the resonance field for the center-mode and a more pronounced decrease for the end-modes. A higher sensitivity to the reduced magnetization at the edges is expected for the end-modes because they are more strongly localized in the area where the magnetization is reduced. The effect of reduced magnetization for the end-modes in the smaller nanomagnets is similar; see Fig. 4 (b). The difference between the experimental spectra and the simulated spectra can be reduced by introducing this small spatial variation in the magnetization and exchange at the edges of the nanomagnets. The influence of the modified edges on the spatial profiles of the center- and end-modes can be seen in the Supplemental Information.

We are careful not to identify reduced edge magnetization as the cause of differences between measured and simulated spectra, as other physical phenomena may produce similar effects. Edge modes in transversely magnetized stripes share characteristics with the end-modes observed here, and the edge modes have been shown to be sensitive side wall angles, edge surface anisotropy and film thickness in addition to magnetization reduction[25].

A comparison of the experimental and simulated spectra shows qualitative agreement with respect to the relative intensity and number of resonances for most cases. However, quantitative agreement between experiment and simulations is limited. Moreover, as can be seen in the Supplemental Information, two SEM images from the same nanomagnet result in two different simulated spectra. This demonstrates that not only details of the nanomagnet shape but also the image used to define the shape of the nanomagnet for the simulations can affect the simulated spectra.

The effect on measurements of nanomagnet arrays due to the variations in the resonance fields for the individual nanomagnets can also be seen in Fig. 5, where we averaged the spectra of 13 of the larger nanomagnets and overlaid two spectra of individual nanomagnets. The linewidth of the averaged center-mode is only slightly increased but the two, individual end-modes cannot be resolved and only one broad peak is visible. This demonstrates again, that the center-mode is less sensitive to nanomagnet to nanomagnet variations. This averaged spectrum also underlines the strength of measurements of individual nanomagnets compared to larger arrays. The individual end-modes would not be resolved, and the linewidth would be determined by a convolution of the distribution of the resonance fields of the end-modes and their relaxation rate.

In conclusion, our comparison between experiment and micromagnetic simulations demonstrates that only qualitative agreement of the spectra can be achieved even when the actual shape of nanomagnets is included in the simulations. The agreement can be improved, when spatial variations of the magnetization and the exchange are included in the simulations. But without prior knowledge of the details of such spatial variations, it is challenging to extract materials parameters by matching experiment to simulations. Finally, this survey of nominally identical nanomagnets, shows how measurements of the building blocks of MRAM devices can provide important insight into device performance. For example, given the difficulty in quantitatively matching the measured dynamics to the simulations, and the fact that even the best process will have variations at the edges, bits that switch by a coherent rotation mechanism will likely have better bit-to-bit variation than those which switch by domain wall propagation.

The data that support the findings of this study are available from the corresponding author upon reasonable request.

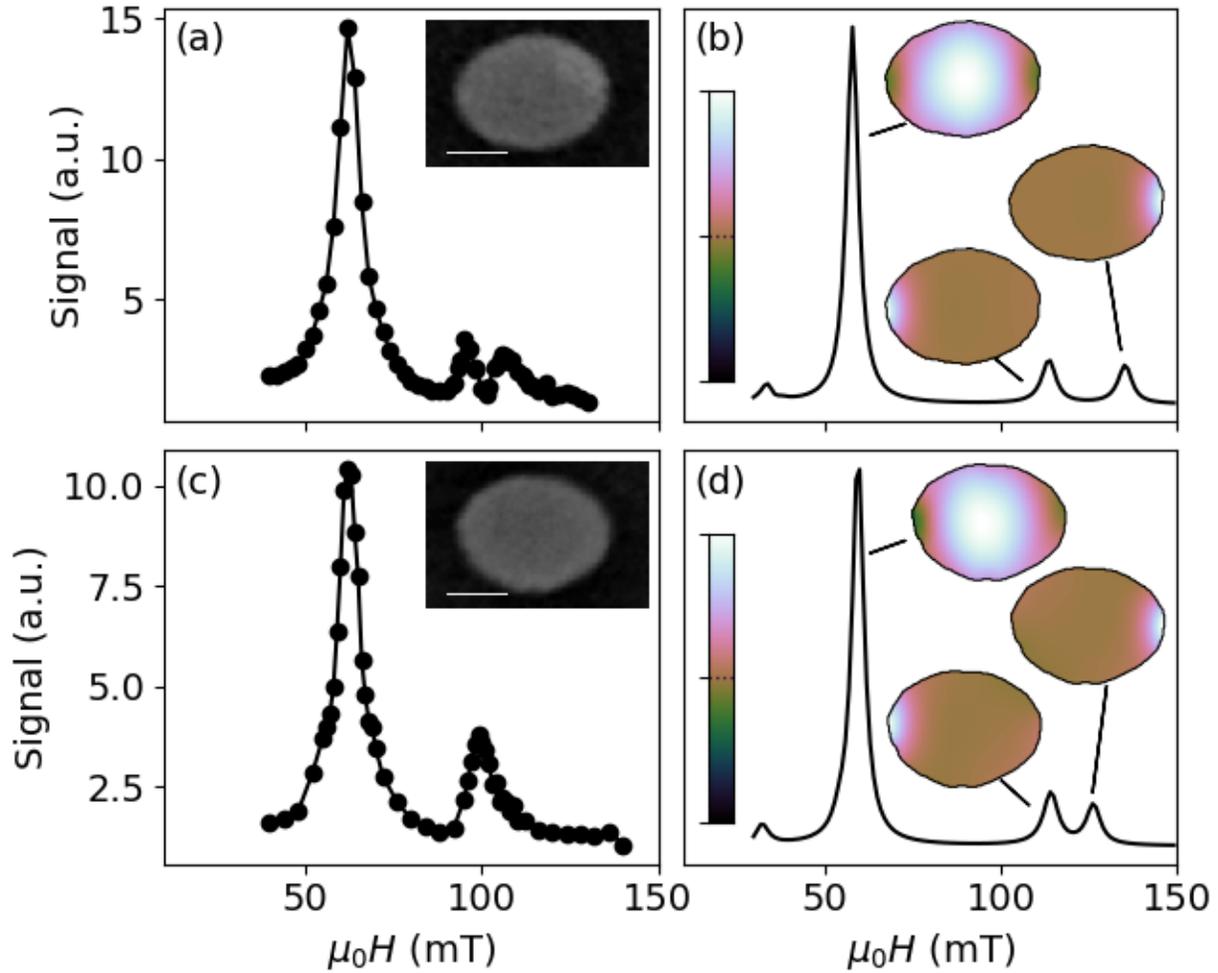

*Figure 1:* Spectra of two 200 nm nanomagnets measured at 8.2 GHz are shown in (a) and (c) together with SEM images of the respective nanomagnets. The corresponding spectra obtained from micromagnetic simulations are shown in (b) and (d), where insets show the modeled mode profiles at the resonance peaks. The strongest peak corresponds to a center-mode, and the two weaker peaks correspond to end-modes. The scale bars are 100 nm. The image outlines show the modeled sample shape as determined from the SEM images.

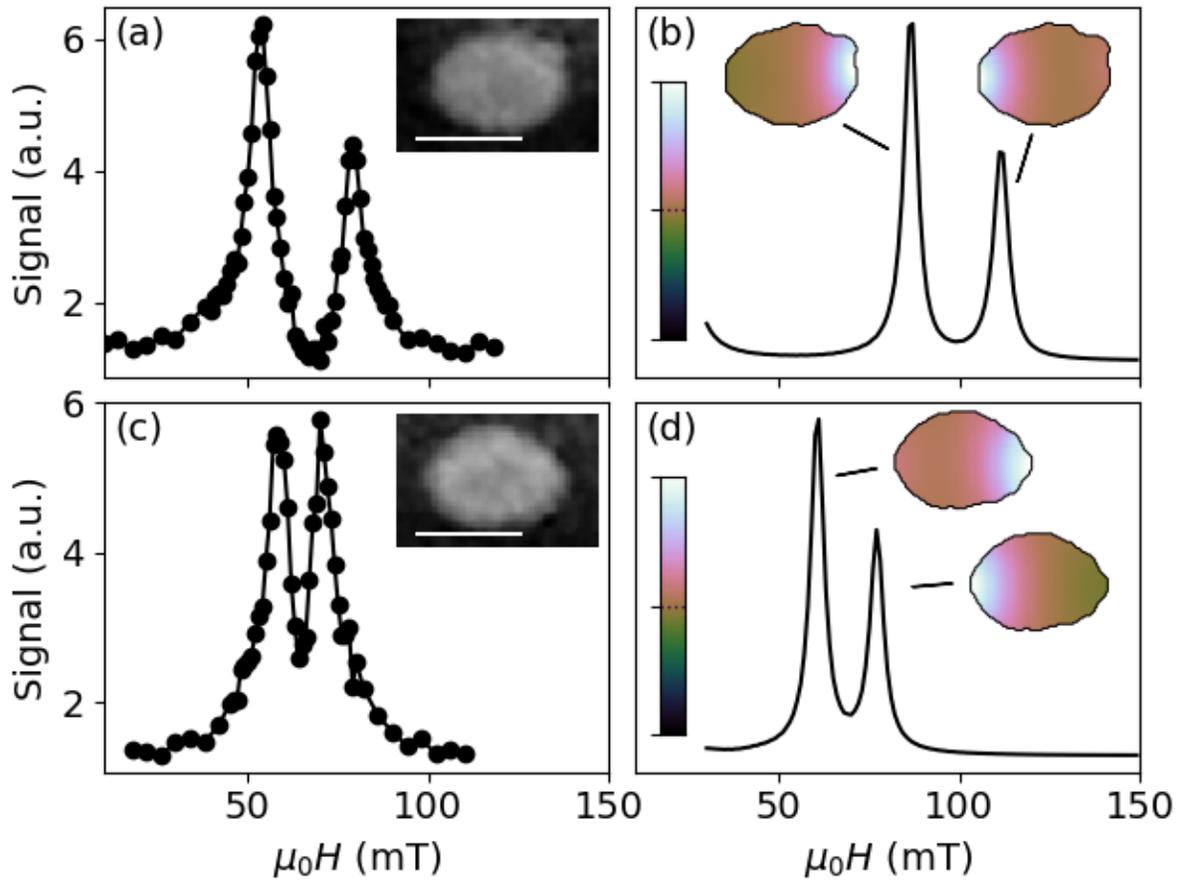

*Figure 2:* Spectra of two 100 nm nanomagnets measured at 8.2 GHz are shown in (a) and (c) together with SEM images of the respective nanomagnets. The corresponding spectra obtained from micromagnetic simulations are shown in (b) and (d), where insets show the modeled mode profiles at the resonance peaks. The peaks observed at this frequency are end-modes. Scale bars are 100 nm, and the image outlines show the modeled sample shape as determined from the SEM images.

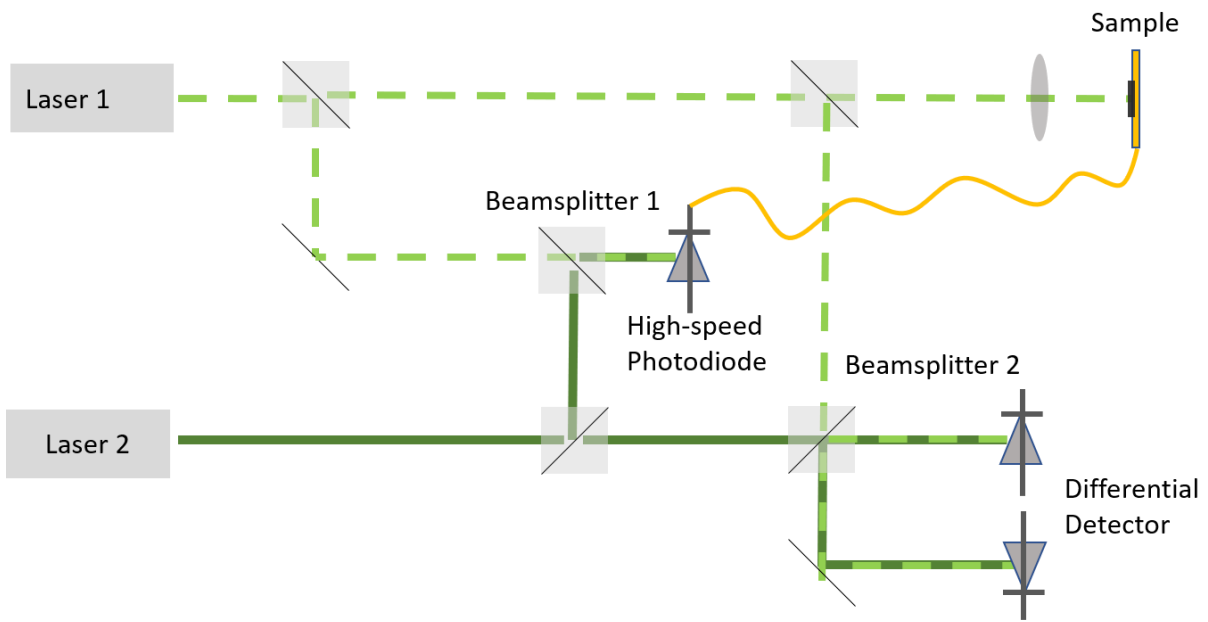

*Figure 3: Sketch of the H-MOMM setup. The beams from the two single frequency lasers are combined at beamsplitter 2 to generate the microwaves. The backreflected laser beam from the sample is mixed with laser 2 at beamsplitter 2, before they are focused onto the differential detector.*

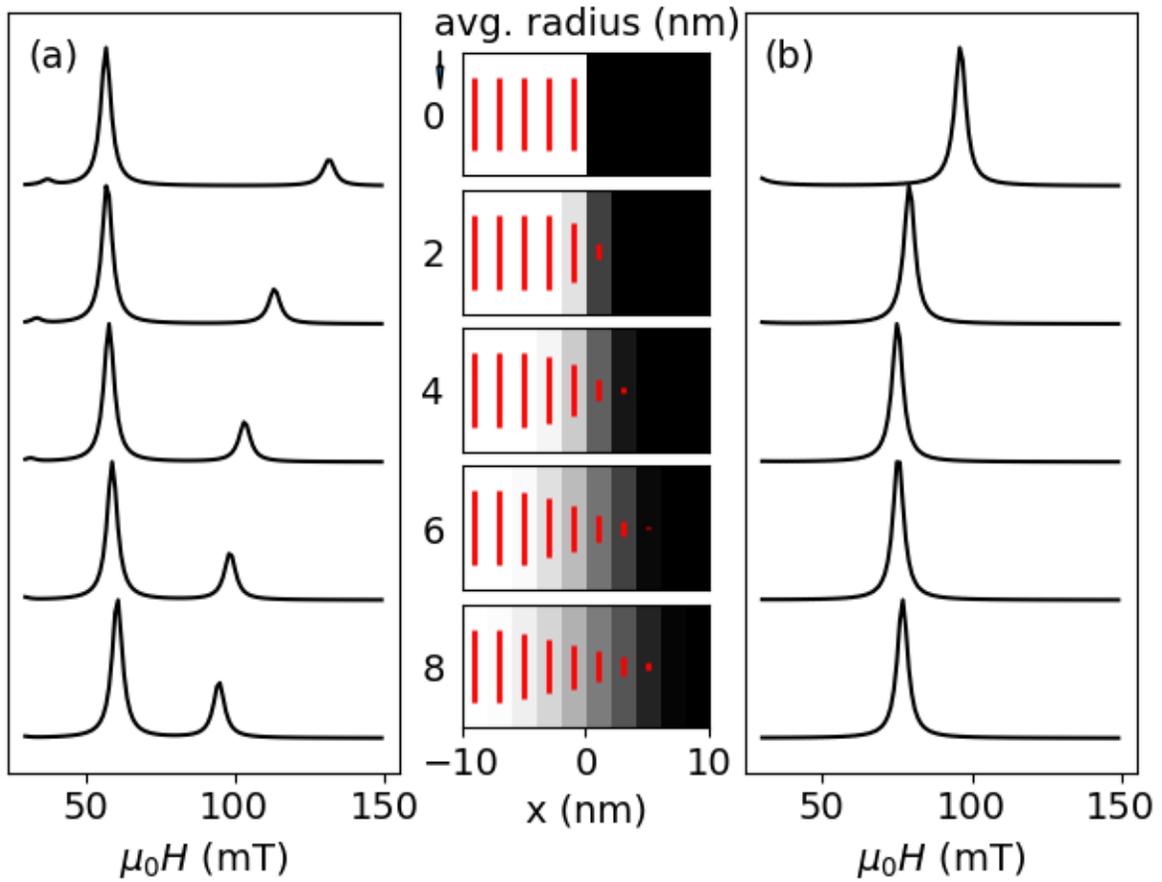

**Figure 4**: Simulated spectra showing the effects of reduced magnetization at the edges. Spectra for symmetric, 262 nm x 190 nm ellipses are plotted in (a) and for 126 nm x 90 nm ellipses in (b). Tapered magnetization profiles were created from ellipse images with moving averages over discs of radius r. The center images show the tapered $M_s$ as red bar lengths and as greyscale. As magnetization near the edge becomes diluted, center-modes (strong peaks in (a)) are largely unaffected while the edge modes (all other peaks) shift to lower fields.

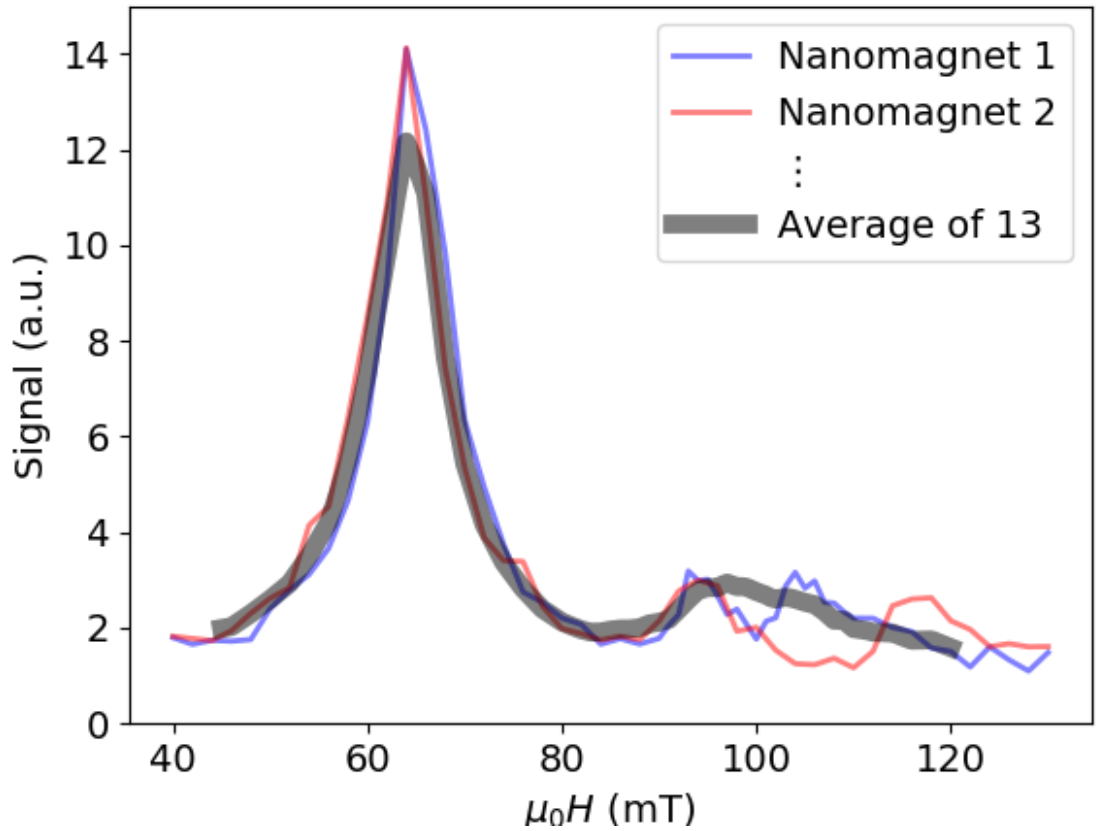

*Figure 5:* *The thick grey line is the average spectra of 13 different nanomagnets. The red and blue traces represent two different spectra of individual nanomagnets.*

**Supplementary Information**

*Comparison of measured and simulated spin-wave mode spectra of magnetic nanostructures*

H. T. Nembach[1,2], R.C. McMichael[3], M.L. Schneider[2], J.M. Shaw[2], T.J. Silva[2]

[1]JILA, University of Colorado, Boulder, Colorado 80309, USA

[2]Quantum Electromagnetics Division, National Institute of Standards and Technology, Boulder, Colorado 80305, USA

[3] Nanoscale Device Characterization Division, National Institute of Standards and Technology, Gaithersburg, Maryland 20899, USA

We carried out H-MOMM measurements and micromagnetic simulations for additional nanomagnets that are not shown in the main article. To test reproducibility of the image-based modeling, two SEM images were taken of most nanomagnets, allowing two independent models of the same magnet. The simulations based on the two images are indicated by the blue and red traces in Figures S1 and S3 below. In general, the resonance fields for the end-modes show a larger image to image variations than the center-modes, which is consistent with the modeled higher sensitivity of end modes to spatial variations of the saturation magnetization, see Fig. 4 in the manuscript. The strong variations of the simulated mode spectra not only from nanomagnet to nanomagnet but also between two simulations based on different images of the same nanomagnet. These differences underscore the difficulty of extracting quantitative information from the comparison between experiment and simulations.

## SI.1 Spin-wave mode spectra

### SI.1a 200 nm Ellipse

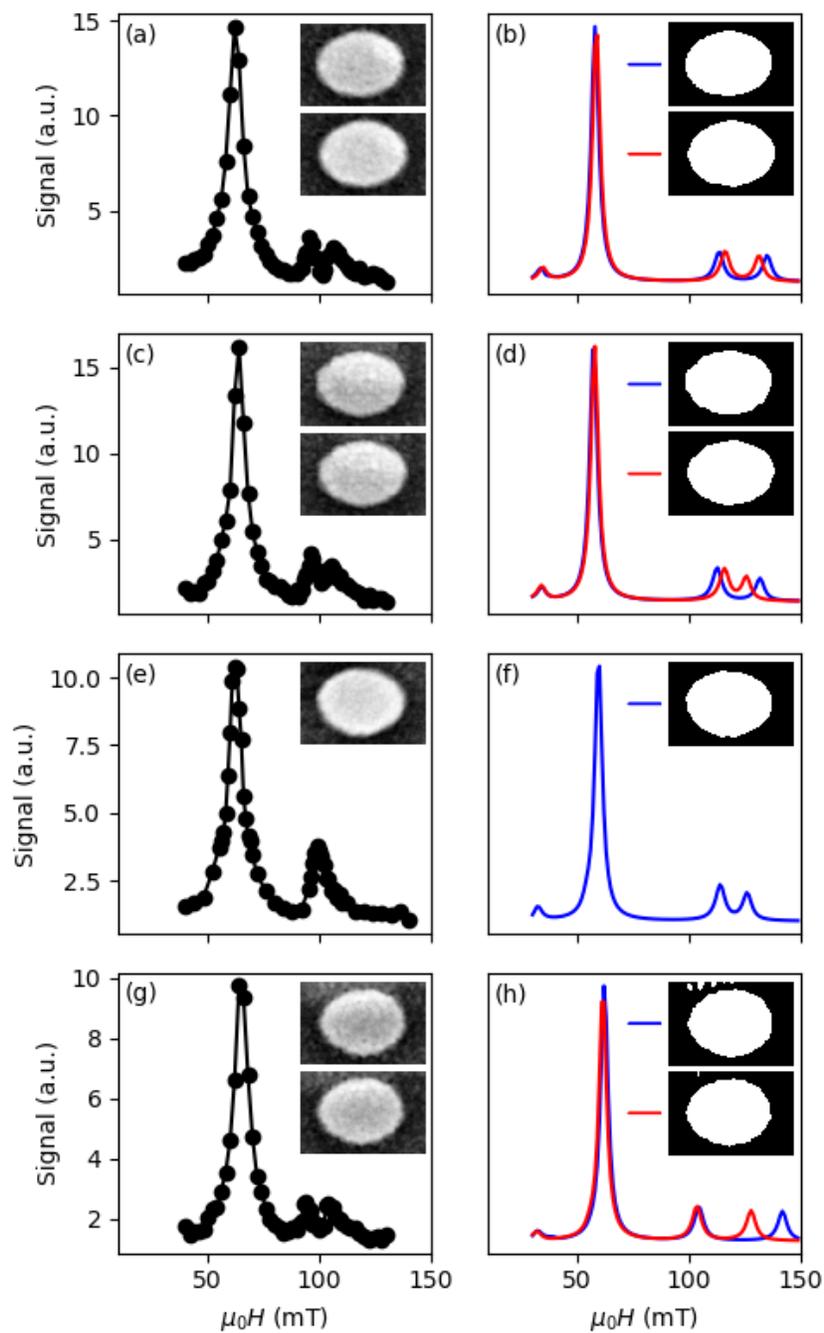

**Figure S1:** Mode spectra for the larger nanomagnets. Two SEM images were taken for most nanomagnets and the results of the micromagnetic simulations are shown in the right column. Fig. S1 (a) and (e) are repeated from Fig 1 (a) and (c) of the main paper.

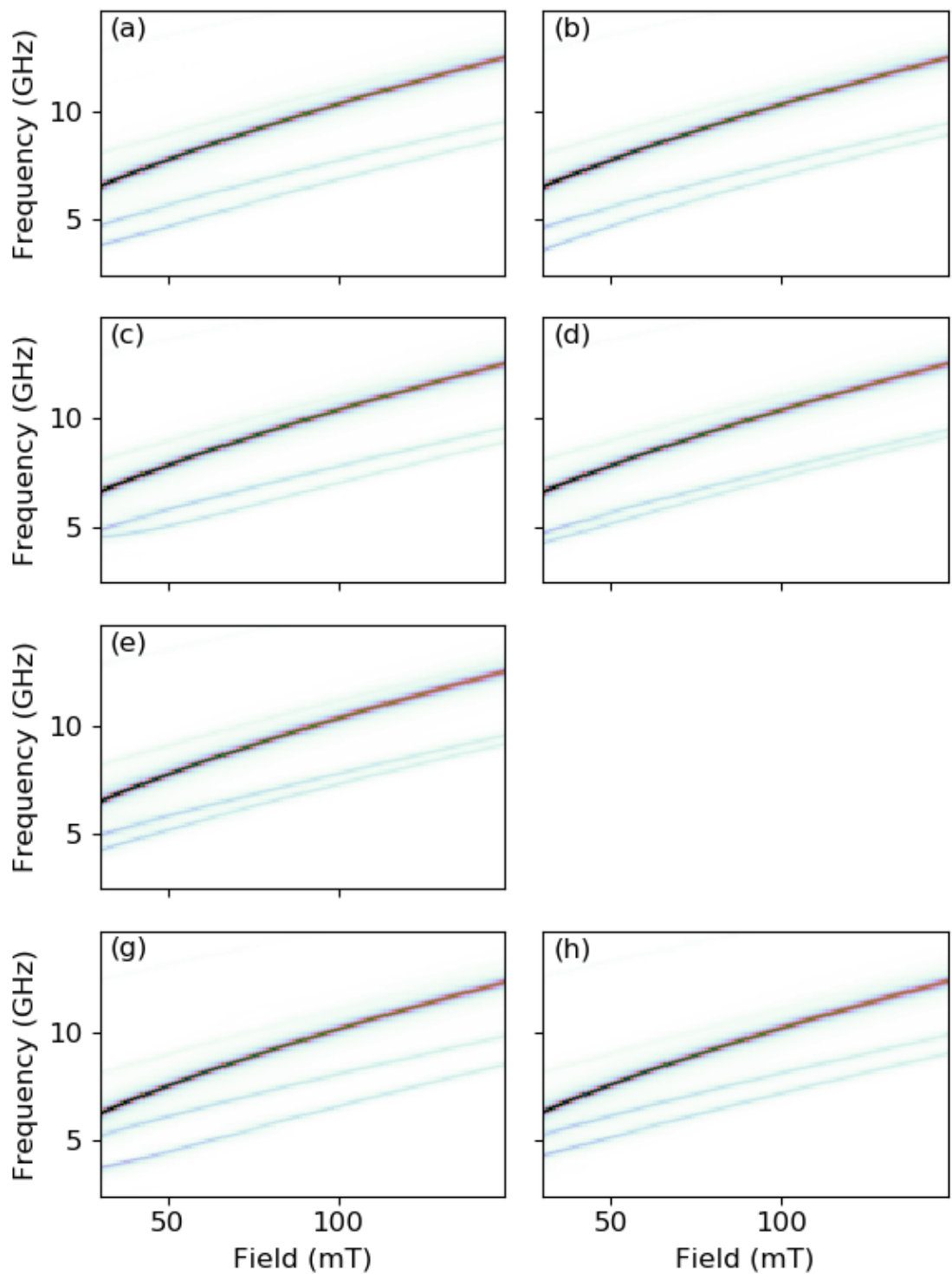

*Figure S2:* Field vs frequency maps modeled for the larger nanomagnets' images. Rows in this figure correspond to the insets in the left column of Fig. S1.

**SI.1b 100 nm ellipses**

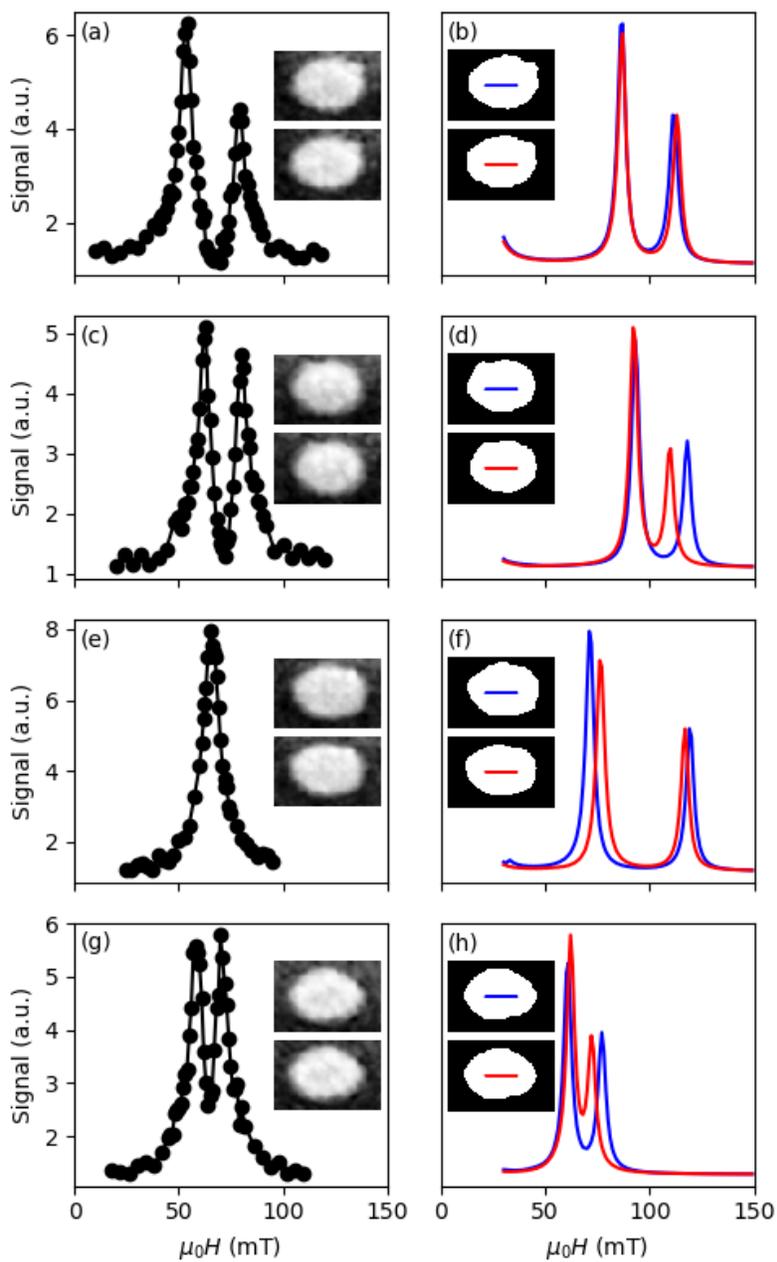

*Figure S3:* Measured (left column) and simulated (right column) spectra of the smaller nanomagnets. Fig. S3 (a) and (g) are repeated from Fig. 2 (a) and (c).

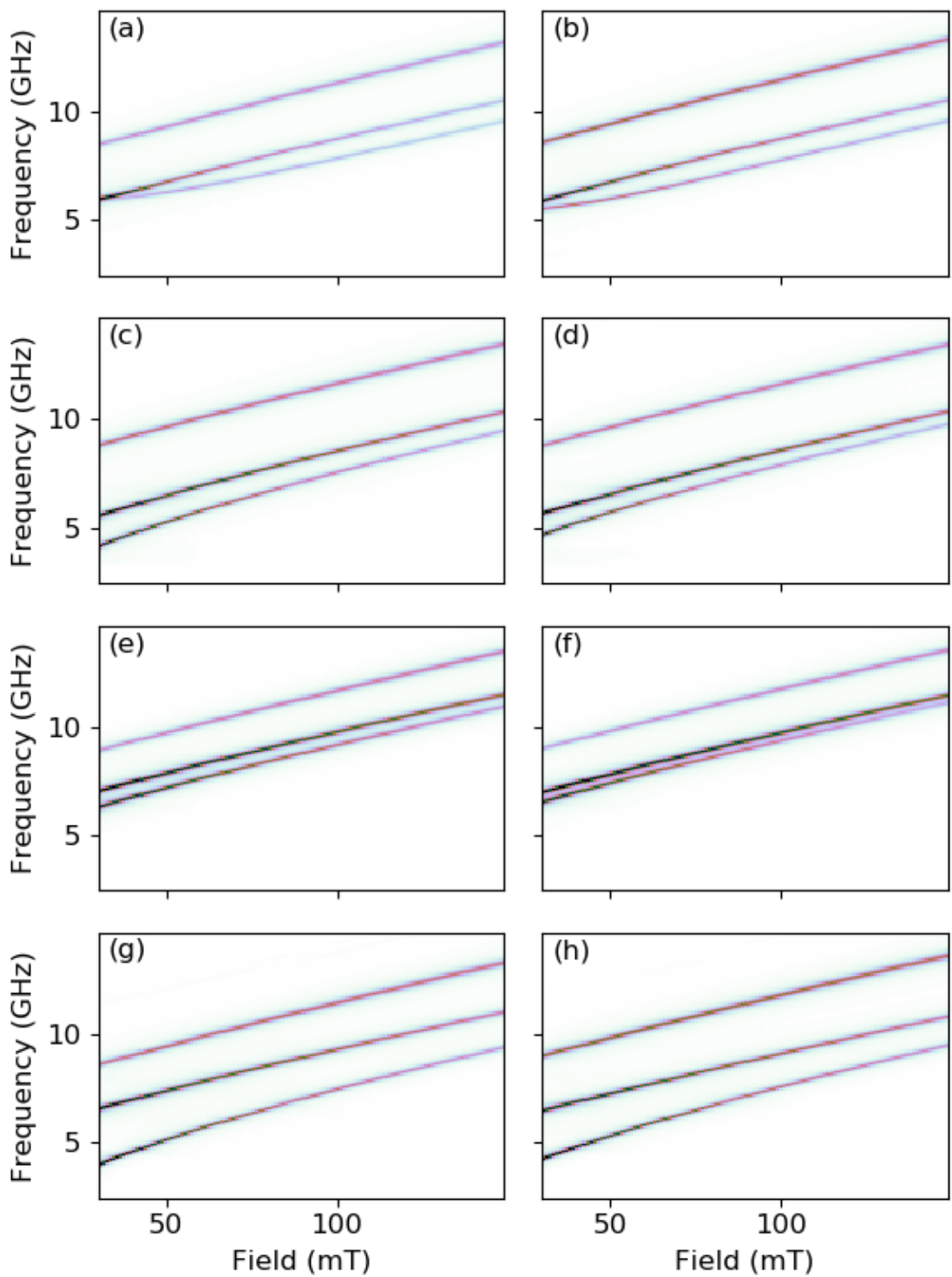

***Figure S4:*** *Frequency vs field maps for the smaller dots. The high-frequency resonance is due to a center mode that lies outside the experimental field range at 8.2 GHz. Rows in this figure correspond to the insets in the left column of Fig. S3.*

## SI.2 Spatial mode maps with edge damage

Fig. 4 in the manuscript shows the dependence of the resonance fields on the averaging radius $r$ for the reduced saturation magnetization and exchange close to the edges. The resonance field for the center-mode of the larger nanomagnets remained unchanged, whereas the resonance field for the end-modes shifts towards lower fields for both nanomagnet sizes. This corresponds to the spatial profiles of the center-mode for the larger nanomagnet remaining mostly unchanged, whereas the end-modes extend further away from the edges with increased $r$. The end-modes for the smaller nanomagnets show a similar change albeit, more pronounced, and the region of opposite phase close to the edges (indicated by black color) for the center-modes increases substantially with $r$.

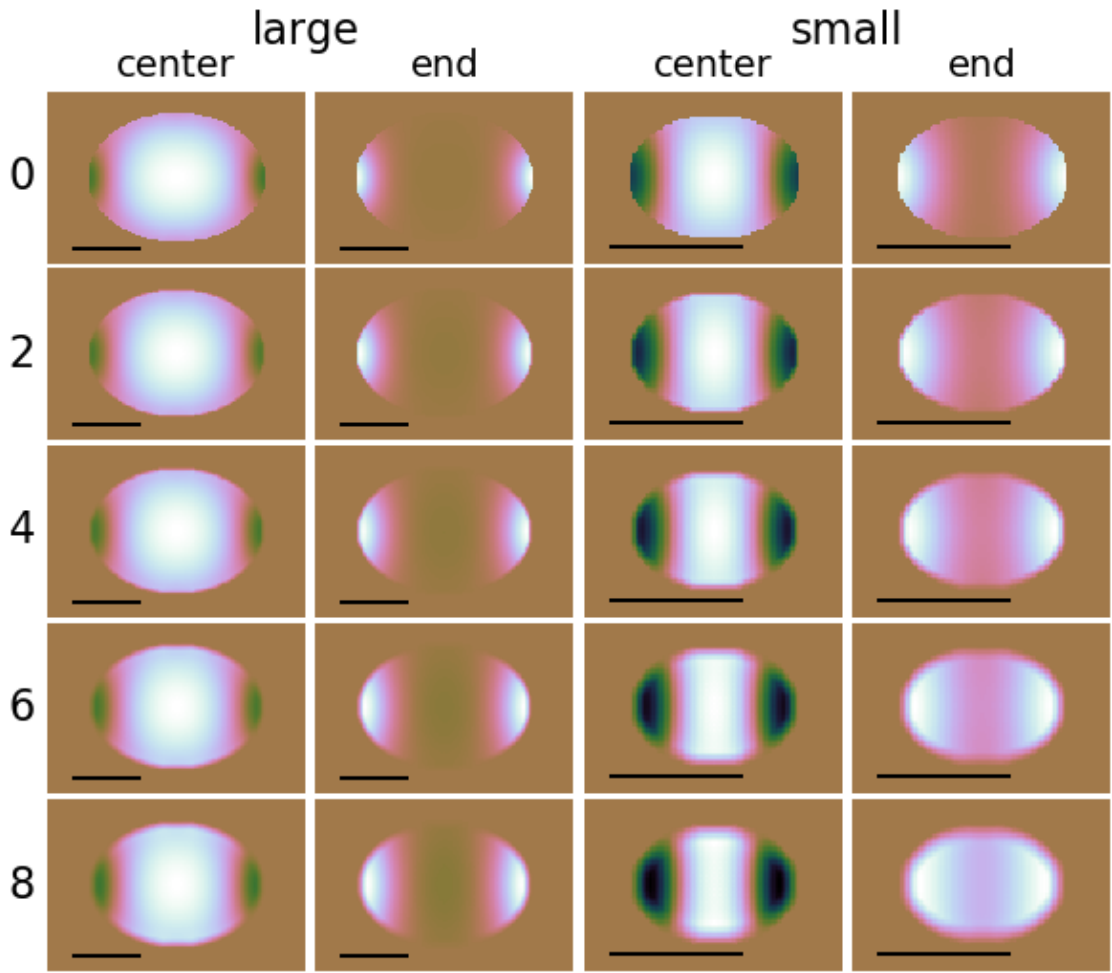

*Figure S5: Dependence of the spatial extend of the spin-wave modes on the saturation magnetization close to the edges. The two left (right) columns are for the large (small) nanomagnets. The number on the left side represents the averaging radius $r$ and the scale bar in each sub-image is 100 nm.*